# Evidence of Collective Flow in p-p Collisions at LHC

Inam-ul Bashir, Riyaz Ahmed Bhat and Saeed Uddin[*],
*Department of Physics, Jamia Millia Islamia (Central University)*
*New Delhi-110025*

## Abstract

The mid-rapidity transverse momentum spectra of hadrons (p, p̄, $K^+$, $K^-$, $K_s^0$, $\phi$, $\Lambda$, $\bar{\Lambda}$, $\Xi^-$, $\overline{\Xi^-}$, ($\Xi^- + \overline{\Xi^-}$), $\Omega$, and $\bar{\Omega}$) and the available rapidity distributions of the strange hadrons ($K_s^0$, ($\Lambda + \bar{\Lambda}$), ($\Xi^- + \overline{\Xi^-}$)) produced in p-p collisions at LHC energy $\sqrt{s_{NN}}$ = 0.9 TeV and $\sqrt{s_{NN}}$ = 7.0 TeV have been studied using a unified statistical thermal freeze-out model. The calculated results are found to be in good agreement with the experimental data. The theoretical fits of the transverse momentum spectra using the model calculations provide the thermal freeze-out conditions in terms of the temperature and collective flow parameters for different hadronic species. The study reveal the presence of significant collective flow and a well defined temperature in the system thus indicating the formation of a thermally equilibrated hydrodynamic system in p-p collisions at LHC. Moreover, the fits to the available experimental rapidity distributions data of strange hadrons show the effect of almost complete transparency in p-p collisions at LHC. The transverse momentum distributions of protons and Kaons produced in p-p collisions at $\sqrt{s_{NN}}$ = 200 GeV and $\sqrt{s_{NN}}$ = 2.76 TeV have also been reproduced successfully. The model incorporates longitudinal as well as a transverse hydrodynamic flow. The contributions from heavier decay resonances have also been taken into account. We have also imposed the criteria of exact strangeness conservation in the system.

*saeed_jmi@yahoo.co.in*

# 1. Introduction

Ultra-relativistic heavy-ion collisions at the Large Hadron Collider (LHC) produce strongly interacting matter under the extreme conditions of temperature and energy density, similar to the conditions prevailing during the first few microseconds of the Universe after the Big Bang [1]. The study of multi particle production in ultra-relativistic heavy-ion collisions also allows us to learn the final state distribution of baryon numbers at the thermo-chemical freeze-out after the collision – initially carried by nucleons only before the nuclear collision [2].

Within the framework of the statistical model it is assumed that initially a fireball, i.e. a hot and dense matter of the partons (quarks and gluons) is formed over an extended region after the collision. The quarks and gluons in the fireball may be nearly free (deconfined) due to the ultra-violet freedom i.e. in a quark gluon plasma (QGP) phase. This fireball undergoes a collective expansion accompanied by further particle production processes through the secondary collisions of quarks and gluons which consequently leads to a decrease in its temperature. Eventually the expansion reaches a point where quarks and gluons start interacting non-perturbatively leading to the confinement of quarks and gluons through the formation of hadrons, i.e. the so called hadronization process. In this hot matter which is in the form of a gas of hadronic resonances at high temperature and density, the hadrons continue to interact thereby producing more hadrons and the bulk matter expands further due to a collective hydrodynamic flow developed in the system. This consequently results in a further drop in the thermal temperature because a certain fraction of the available thermal energy is converted into directed (collective hydrodynamic) flow energy. As the mean free paths for different hadrons, due to expansion increases, the process of decoupling of the hadrons from the rest of the system takes place and the hadron spectra are frozen in time. The hadrons with smaller cross-sections stop interacting with the surrounding matter earlier and hence decouple earlier. Hence a so called *sequential* thermal/kinetic freeze-out of different hadronic species occurs. Following this, the hadrons freely stream out to the detectors. The freeze-out conditions of a given hadronic specie are thus directly reflected in its transverse momentum and rapidity spectra [3].

Within the framework of the statistical model [4] the measured particle ratios can be used to ascertain the system temperature and the baryonic chemical potential, $\mu_B$, at the final freeze-out i.e. at the end of the evolution of the hadronic gas phase. The statistical model thus assumes that the system is in thermal and chemical equilibrium at this stage. The system at freeze-out can be described in terms of a nearly free gas of various hadronic resonances (HRG). The above assumptions are valid with or without the formation of a QGP at the initial stage. It is believed that the produced hadrons also carry information about the collision dynamics and the subsequent space-time evolution of the system. Hence precise measurements of the transverse momentum distributions of identified hadrons along with the rapidity spectra are essential for the understanding of the dynamics and properties of the created matter up to the final freeze-out [5]. The transverse momentum distributions are believed to be encoded with the information about the collective transverse and longitudinal expansions and the thermal temperature at freeze-out.

The particle production in p-p collisions are very important as these can serve as a baseline for understanding the particle production mechanism and extraction of the signals of QGP formation in heavy ion collisions [6].

The value of chemical potential is always lower in p-p collisions than in heavy ion collisions due to the lower stopping power in p-p collisions [7]. As one goes to LHC energies, the stopping reduces much further giving rise to nearly zero net baryon density at mid-rapidity and thus the value of the chemical potential at mid-rapidity essentially reduces to zero. Thus at LHC, we believe the p-p collisions to be completely transparent.

The p-p collisions at lower energies were successfully described in the past by using statistical hadronization model [8, 9]. The same kind of analysis has been performed on the p-p results at LHC energy of 0.9 TeV [10]. Naively, the p-p collisions are not expected to form QGP or a system with collective hydrodynamic effects. An absence of radial flow in p-p collisions at $\sqrt{s_{NN}}$ = 200 GeV and 540 GeV was found in a recent work [11]. However, there have been speculations [12–15] about the possibility of the formation of such a system but of smaller size in the p-p collisions. The occurrence of the high energy density events in high multiplicity p-$\bar{p}$ collisions [16, 17] at CERN-SPS motivated searches for hadronic deconfinement in these collisions at $\sqrt{s_{NN}}$ = 0.54 TeV at SPS [12] and at $\sqrt{s_{NN}}$ = 1.8 TeV [13, 14] at the Tevatron,

Fermilab. A common radial flow velocity for meson and anti-baryon found from the analysis of the transverse momentum data of the Tevatron [13] had been attributed to as an evidence for collectivity due to the formation of QGP [18].

Keeping in view the above facts, we in our present analysis will address the collective effect signatures in the p-p collisions at LHC, particularly in terms of transverse and longitudinal flows while attempting to reproduce the transverse momentum and rapidity distributions of hadrons produced in p-p collisions at different LHC energies.

## 2. Model

Though the details of the model used here can be found elsewhere [5] however for the sake of convenience we will briefly describe our model here. In a statistical hydrodynamic description taking into account the flow in the transverse and longitudinal directions, the final state particles will leave the hadronic resonance gas (HRG) at the time of freeze-out. The momentum distributions of hadrons, emitted from within an expanding fireball, assumed to be in the state of local thermal equilibrium, are characterized by the Lorentz-invariant Cooper-Frye formula [19]

$$E \frac{d^3 n}{d^3 P} = \frac{g}{(2\pi)^3} \int f\left(\frac{p^\mu u^\mu}{T}, \lambda\right) p^\mu d\Sigma_\mu, \quad (1)$$

Where $\Sigma_f$ represents a 3-dimensional freeze-out hyper-surface, g is the degeneracy factor of a given hadronic specie in the expanding relativistic hadronic gas and $\lambda$ is the fugacity of the given hadronic specie. In recent works [20,21] it has been clearly shown that there is a strong evidence of increasing baryon chemical potential, $\mu_B$ along the collision axis at RHIC. This effect, which is an outcome of the nuclear transparency effect [5] is incorporated by writing [21,22]

$$\mu_B = a + b y_0^2 \quad (2)$$

Where $y_0$ ($=\xi z$) is the rapidity of the expanding hadronic fluid element along the beam axis (z-axis).

The transverse velocity component of the hadronic fireball, $\beta_T$ is assumed to vary with the transverse coordinate $r$ in accordance with the Blast Wave model as [23]

$$\beta_T(r) = \beta_T^s \left(\frac{r}{R}\right)^n \quad (3)$$

Where $n$ is an index which determines the profile of $\beta_T(r)$ and $\beta_T^s$ is the hadronic fluid *surface* transverse expansion velocity and is fixed in the model by using the following parameterization [5]

$$\beta_T^s = \beta_T^0 \sqrt{1 - \beta_z^2} \qquad (4)$$

The transverse radius of fireball $R$ is parameterized as [5, 24]

$$R = r_0 \, exp\left(-\frac{z^2}{\sigma^2}\right) \qquad (5)$$

Where the parameter $r_0$ fixes the *transverse* size of the expanding hadronic matter at the freeze-out, along with $\sigma$, for different values of the longitudinal i.e. the *z*-coordinate [5]. The contributions of various heavier hadronic resonances [21, 25], which decay after the freeze-out has occurred, are also taken into account in our analysis. In our study we have included only the two body decay channels which dominate the decay contributions to the hadrons considered. Also if the decay channels are classified as dominant, large seen or possibly seen, we take into account the dominant channel only. If two or more than two channels are described as equally important, we take all of them with the same weight. We have considered the baryonic and mesonic resonances having masses up to 2 GeV. We also impose the criteria of exact strangeness conservation.

## 3. Rapidity Spectra

In Figure 1, we have shown the rapidity distributions of some strange particles like $K_s^0$, ($\Lambda + \bar{\Lambda}$) and ($\Xi^- + \overline{\Xi^-}$) at $\sqrt{s_{NN}}$ = 0.9 TeV and 7.0 TeV. The available data is taken from CMS experiment at CERN LHC [26] and is shown by colored filled shapes. The best fit is obtained by minimizing the distribution of $\chi^2$ given by [27], (9)

$$\chi^2 = \sum_i \frac{(R_i^{exp} - R_i^{theor})^2}{\epsilon_i^2} \qquad (6)$$

We have taken only statistical errors into consideration. The $\chi^2/DoF$ for fitting the rapidity spectra are minimized with respect to the variables $a$, $b$, $\sigma$ and $r_0$, whereas the values of T, n and $\beta_T^0$ are first obtained by fitting the corresponding $p_T$ distributions. The $p_T$ distributions are not affected by the values of $a$, $b$, $\sigma$ and $r_0$, instead these parameters have significant effect on the rapidity distribution shapes. The fit parameters obtained from the rapidity distributions of the three experimental data set at $\sqrt{s_{NN}}$ = 0.9 TeV and 7.0 TeV are given in Table 1 below.

| $\sqrt{s_{NN}}$ | Particle | $a$ (MeV) | $b$ (MeV) | $\sigma$ (fm) | $r_0$ |
|---|---|---|---|---|---|
| 0.9TeV | $K_s^0$ | 1.88 | 3.70 | 5.40 | 5.70 |
| | ($\Lambda + \bar{\Lambda}$) | 1.35 | 3.55 | 4.70 | 5.70 |
| | $\Xi^- + \overline{\Xi^-}$ | 1.0 | 3.45 | 4.40 | 5.40 |
| 7.0TeV | $K_s^0$ | 1.25 | 7.13 | 5.36 | 5.75 |
| | ($\Lambda + \bar{\Lambda}$) | 0.90 | 6.45 | 4.75 | 5.70 |
| | $\Xi^- + \overline{\Xi^-}$ | 0.70 | 5.52 | 4.40 | 5.70 |

**Table.1** Values of a, b, $\sigma$ and $r_0$ obtained from fitting the rapidity distributions of $K^0_s$, $(\Lambda + \overline{\Lambda})$ and $(\Xi^- + \overline{\Xi^-})$, respectively, at $\sqrt{s_{NN}} = 0.9$ TeV and 7.0 TeV.

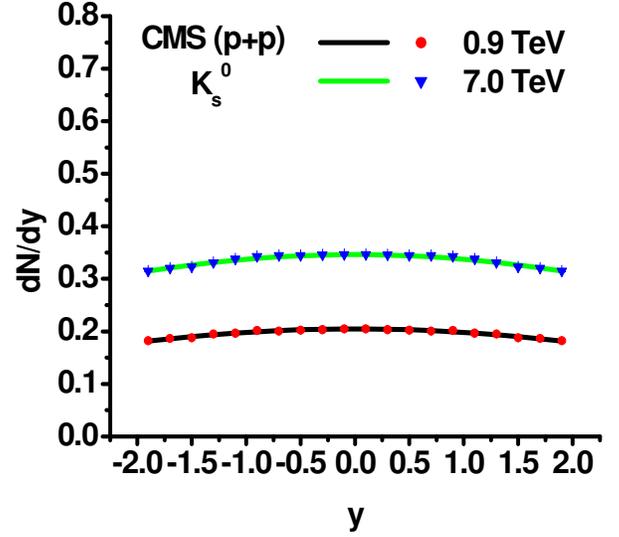

It is evident from Table 1 that the value of the baryonic chemical potential, as given by eq. 2, approaches to zero in these experiments in the rapidity range of $0 \pm 2$ units. At $\sqrt{s_{NN}} = 7.0$ TeV a smaller value of *a* and a larger value of *b* indicates a higher degree of nuclear transparency [5]. However, on the overall basis it can be said that these LHC experiments involving p-p collisions give a clear indication of the existence of a nearly baryon free matter owing to a high degree of nuclear transparency effect. Another evidence for this transparency also comes from [28] where the measured mid-rapidity anti-baryon to baryon ratio is found to be nearly equal to unity at various LHC energies. This fact is also supported by the nearly flat rapidity distributions in Figure 1.

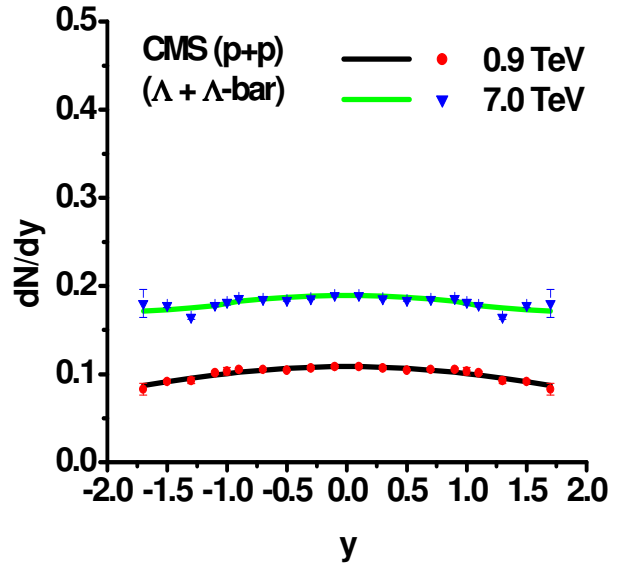

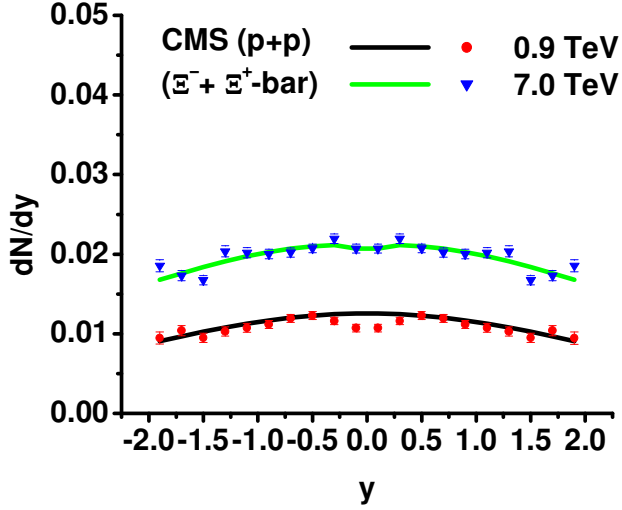

**Figure:1.** Rapidity distribution of $K_s^0$, $(\Lambda+ \overline{\Lambda})$, $\overline{\Lambda}/\Lambda$ and $(\Xi^- +\overline{\Xi^+})$ at $\sqrt{s_{NN}}$=0.9 TeV and 7.0 TeV

## 4. Transverse momentum spectra at $\sqrt{s_{NN}} = 0.9$ TeV

The transverse momentum distributions of various hadrons produced in p-p collisions at $\sqrt{s_{NN}} = 0.9$ TeV are fitted. We find that the model calculations (shown by black solid curves) agree quite well with the experimental data (shown by red filled circles) taken from the ALICE Collaboration at $\sqrt{s_{NN}} = 0.9$ TeV [29]. The values of the parameters T, n and $\beta_T^0$ at freeze-out are obtained through a best fit to a given hadron's transverse momentum spectrum. These values are then used to fit the rapidity data and determine the values of *a, b, σ* and $r_0$ from the available rapidity spectra of the hadrons.

These values are given in Table 1. The flow velocity profile index *n* varies from 1.0 to 1.20 for different cases. The value of c=1fm$^{-1}$ is fixed for all the hadrons studied in this paper. The available error bars here represent the sum of statistical and systematic uncertainties. In the Figure 2, we have shown the transverse momentum spectra of protons and antiprotons at $\sqrt{s_{NN}} = 0.9$ TeV. The model curves cross virtually all data points within the error bars. The values of the freeze-out parameters of the hadrons obtained from their transverse momentum spectra, along with their minimum $\chi^2/dof$ are shown in Table 2. The similar values of the freeze-out parameters for protons and antiprotons indicate a near simultaneous freeze-out of these particles from the dense hadronic medium.

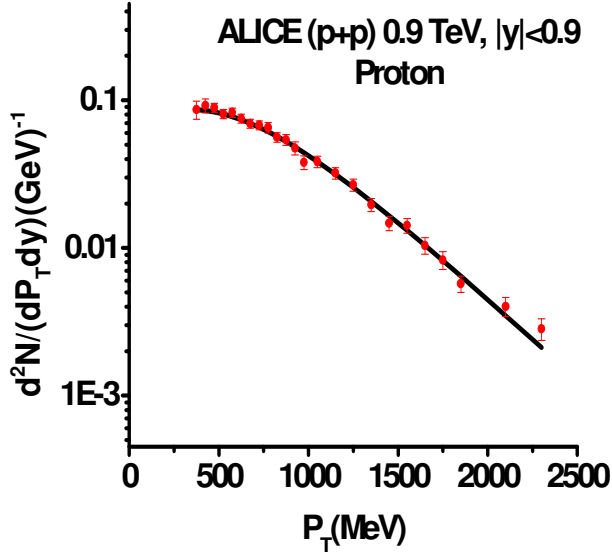

| | | | | |
|---|---|---|---|---|
| $\bar{p}$ | 173.0 | 0.56 | 1.11 | 1.0 |
| $\phi$ | 175.0 | 0.52 | 1.0 | 0.25 |
| $\Lambda$ | 175.0 | 0.51 | 1.02 | 0.70 |
| $\bar{\Lambda}$ | 176.0 | 0.51 | 1.0 | 0.41 |
| $(\Xi^- + \overline{\Xi^-})$ | 176.0 | 0.49 | 1.02 | 0.91 |

**Table.2. Freeze-out parameters of various hadron along with their corresponding $\chi^2/dof$, produced at $\sqrt{s_{NN}} = 0.9$ TeV.**

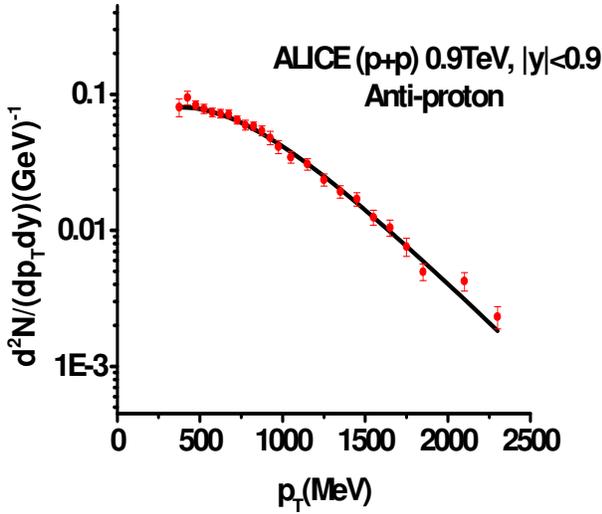

**Figure:2. Transverse momentum spectra of protons p and antiprotons $\bar{p}$.**

| Particle | $T$ (MeV) | $\beta_T^0$ | $n$ | $\chi^2/dof$ |
|---|---|---|---|---|
| $K^+$ | 173.0 | 0.58 | 1.20 | 1.18 |
| $K^-$ | 174.0 | 0.58 | 1.20 | 0.76 |
| $K_S^0$ | 174.0 | 0.55 | 1.04 | 0.63 |
| $p$ | 172.0 | 0.56 | 1.10 | 0.65 |

The transverse momentum spectra for $K^+$ and $K^-$ are shown in Figure 3. We observe a good agreement of our model calculations with the experimental data upto $p_T$=2.0 GeV. At larger values of $p_T$, where hard processes are expected to contribute, the model falls below the data for $K^+$ and $K^-$. Also the model predictions for these particles are the same because of the vanishing chemical potential at mid-rapidity. The similar freeze-out parameters obtained for Kaons and anti-Kaons indicate a simultaneous freeze-out of these particles from the dense hadronic medium.

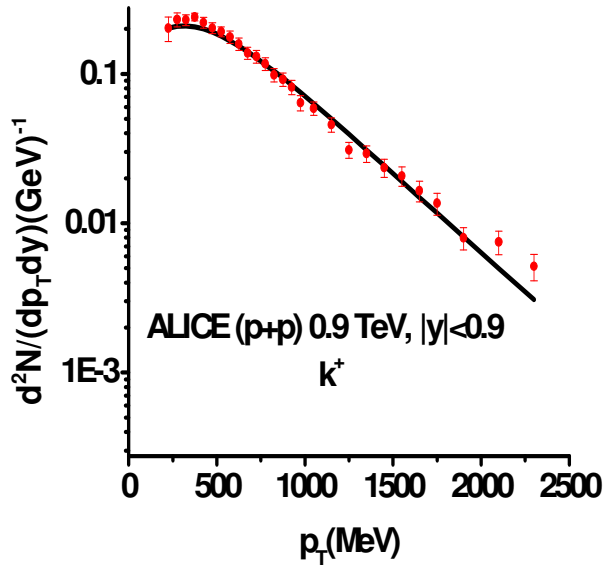

very good agreement between the model calculations and the experimental data points for the $K_s^0$ case. Also the predicted spectrum of the ϕ mesons agrees well with the experimental data. The φ meson serves as a very good "thermometer" of the system. This is because its interaction with the hadronic environment is negligible. Moreover, it receive almost no contribution from resonance decays, hence its spectrum directly reflects the thermal and hydrodynamical conditions at freeze-out.

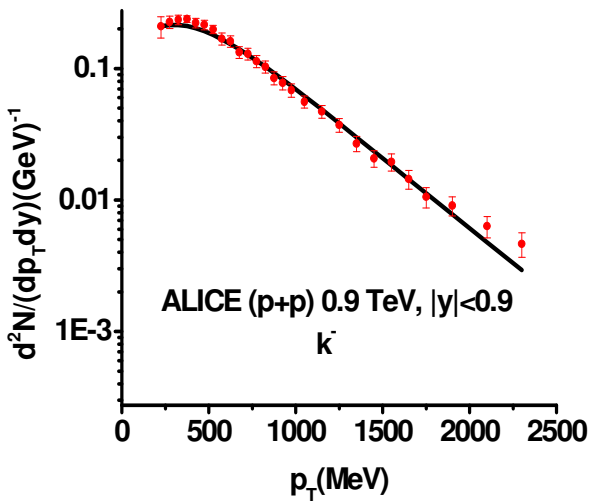
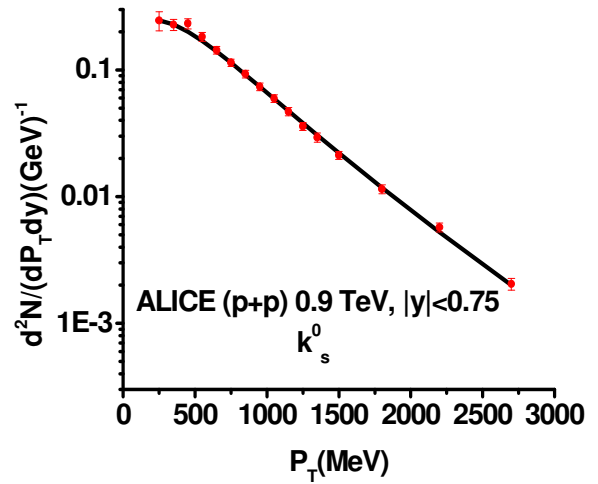

**Figure:3. Transverse momentum spectra of $K^+$ and $K^-$.**

The transverse momentum spectra of $K_s^0$ and ϕ meson are shown in Figure 4. We observe a

9/17

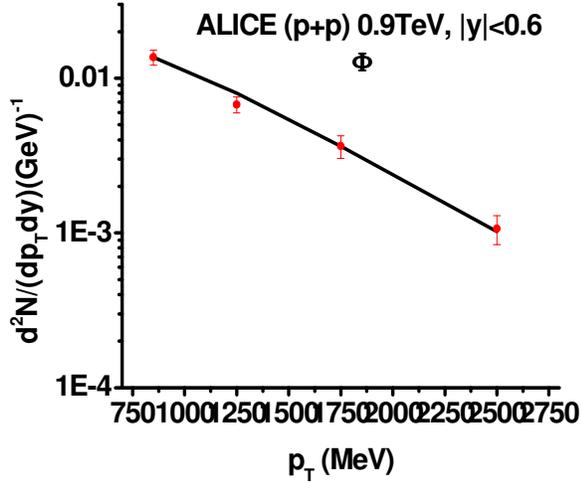

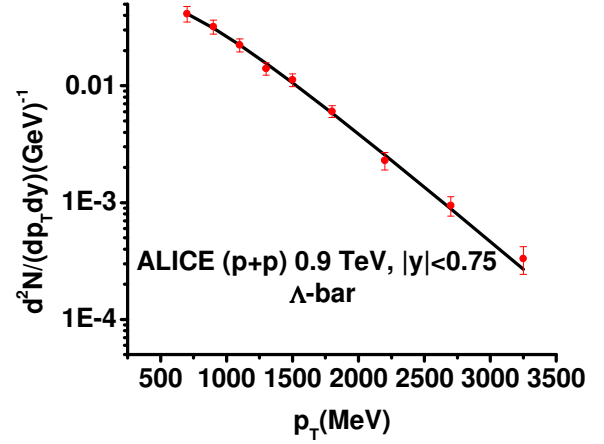

**Figure:4.** Transverse momentum spectra of $K^0_s$ and $\phi$ meson.

**Figure:5.** Transverse momentum spectra of lambda $\Lambda$ and anti-lambda $\overline{\Lambda}$.

The transverse momentum spectra of $\Lambda$ and $\overline{\Lambda}$ are shown in Figure 5. The model curves are found to cross virtually all data points within the error bars. These particles are again found to freeze-out simultaneously as indicated by their similar freeze-out conditions.

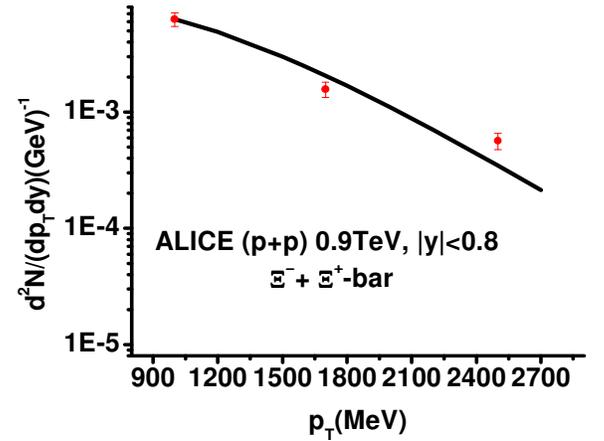

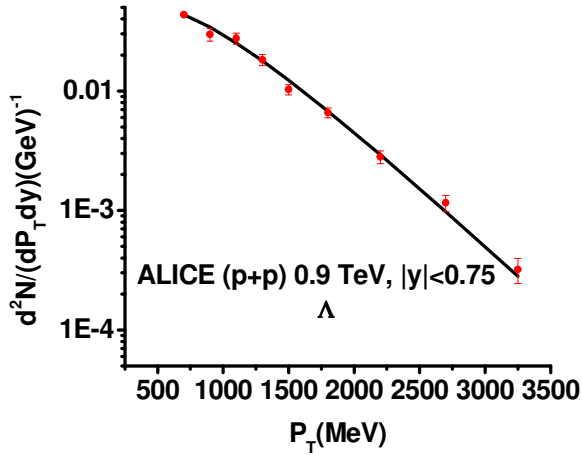

**Figure:6** The $p_T$ spectra of $(\Xi^- + \overline{\Xi^-})$.

It is clear that the heavier particles have somewhat more flattened transverse momentum distribution as compared to lighter particles thereby exhibiting a larger apparent temperature. This is developed by their early thermo-chemical freeze-out in the system. The early decoupling of multi-strange hyperons also result due to their lower interaction cross section with the surrounding hadronic matter of the fireball formed out of an ultra-relativistic proton-proton collision.

Further we have compared our theoretical results with the calculations from the PYTHIA event generator [30] using two different tunes indicated by colored curves in Figure 7, for the case of $\Lambda$ and $K_s^0$. It is seen that the PYTHIA curves underestimates the experimental data pattern in the case of $\Lambda$ and $K_s^0$ and they fall below the data points especially at high $p_T$. In the case of PHOJET calculations a similar behavior like PYTHIA is again exhibited. Similar deviations from the experimental data are observed for the case of $\phi$ meson and ($\Xi^-$ + $\overline{\Xi^-}$) [30] (not shown here). In comparison to PYTHIA and PHOJET calculations, our model successfully reproduces the spectra of these particles over the whole $p_T$ range.

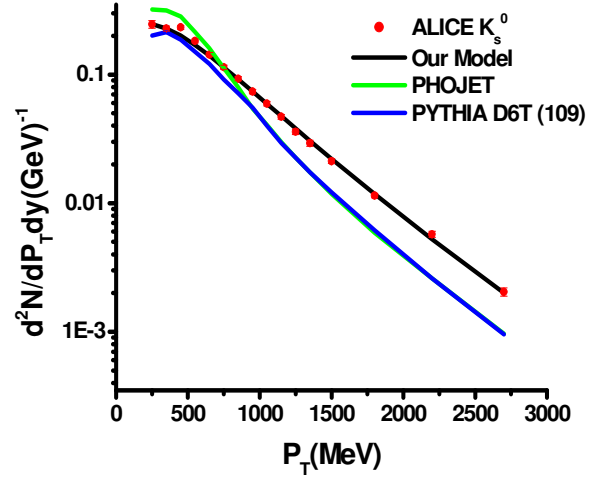

**Figure:7. Comparison of our model results with PHOJET and PYTHIA D6T (109) calculations.**

## 5. Transverse momentum spectra at $\sqrt{s_{NN}}$ = 7.0 TeV.

In figures 8-11 the transverse momentum spectra of the hadrons i.e. p, $\bar{p}$, $K^+$, $K^-$, $\phi$, $\Xi^-$, $\overline{\Xi^-}$, $\Omega$ and $\overline{\Omega}$ produced in p-p collisions at $\sqrt{s_{NN}}$ = 7.0 TeV are shown. The experimental data [31] is again fitted quite well with our model calculations. The flow velocity profile index $n$ varies from 1.02 to 1.17. The error bars here represent the sum of statistical and systematic uncertainties. The various freeze-out conditions obtained from the $p_T$ spectra of these hadrons are given in Table 3.

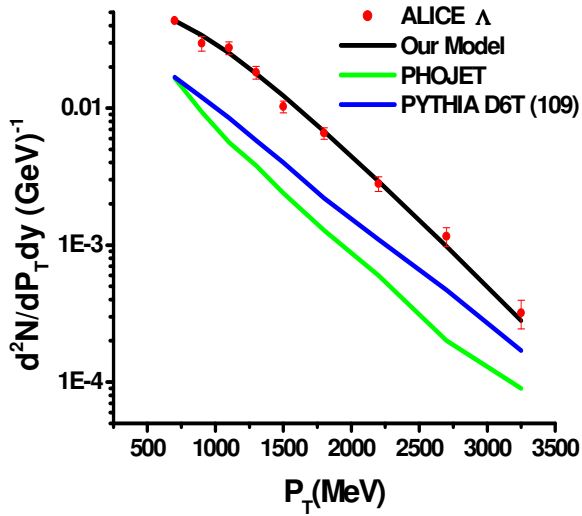

| Particle | $T$ (MeV) | $\beta_T^0$ | n | $\chi^2$/dof |
|---|---|---|---|---|
| $K^+$ | 172 | 0.79 | 1.17 | 0.50 |
| $K^-$ | 173 | 0.79 | 1.17 | 0.53 |
| $p$ | 174 | 0.83 | 1.10 | 1.85 |
| $\bar{p}$ | 175 | 0.83 | 1.10 | 2.10 |
| $\phi$ | 175 | 0.79 | 1.08 | 0.15 |
| $\Xi^-$ | 176 | 0.71 | 1.06 | 0.72 |
| $\overline{\Xi^-}$ | 177 | 0.71 | 1.05 | 0.70 |
| $\Omega$ | 177 | 0.70 | 1.02 | 0.87 |
| $\overline{\Omega}$ | 178 | 0.70 | 1.02 | 0.47 |

**Table 3.** Freeze-out conditions of hadrons produced at $\sqrt{s_{NN}} = 7.0$ TeV with corresponding $\chi^2/dof$.

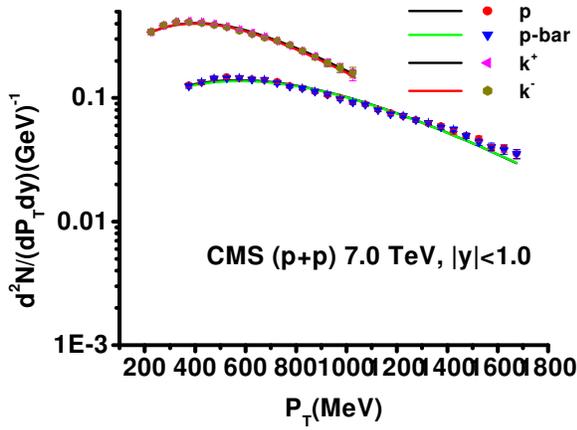

**Figure:8.** Transverse momentum spectra of p, $\bar{p}$, K$^+$ and K$^-$ at $\sqrt{s_{NN}} = 7.0$ TeV.

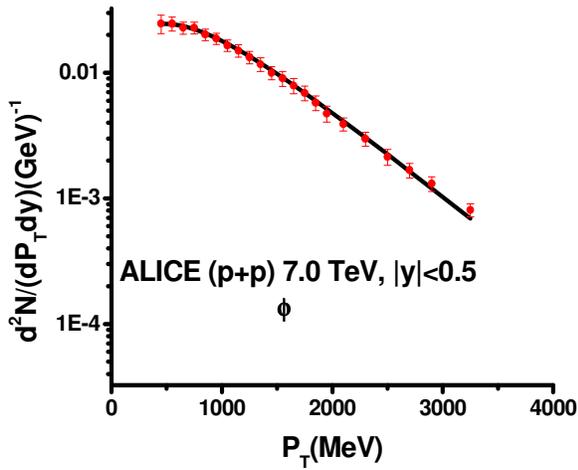

**Figure:9.** The $p_T$ spectrum of $\phi$- meson.

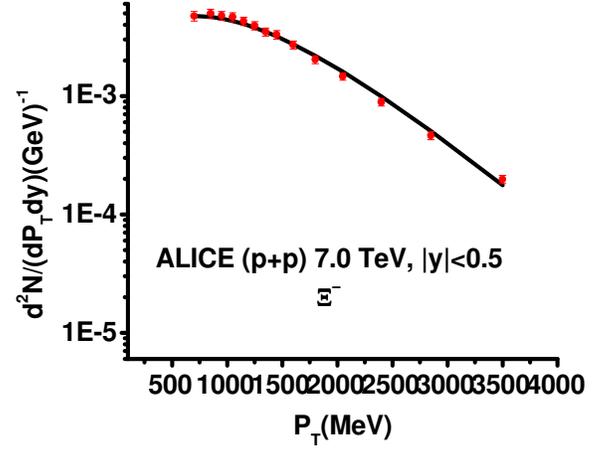

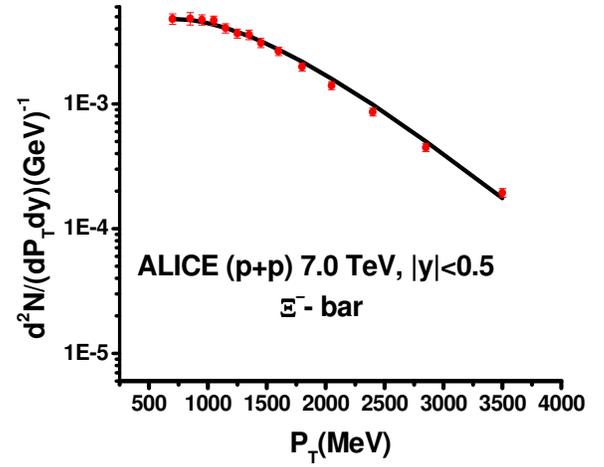

**Figure:10.** The $p_T$ spectra of $\Xi^-$ and $\overline{\Xi^-}$.

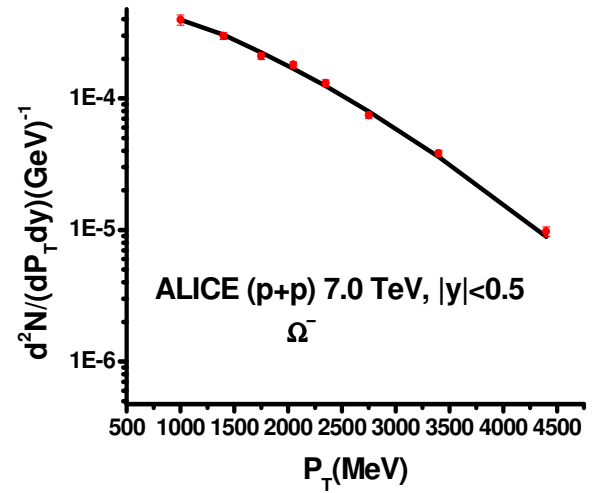

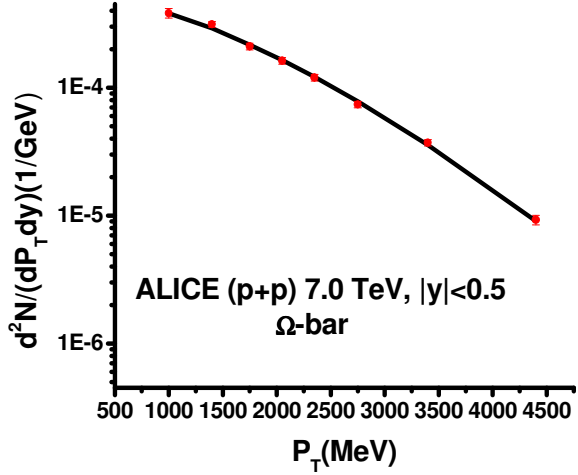

**Figure:11.** The $p_T$ spectra of $\Omega$, and $\bar{\Omega}$.

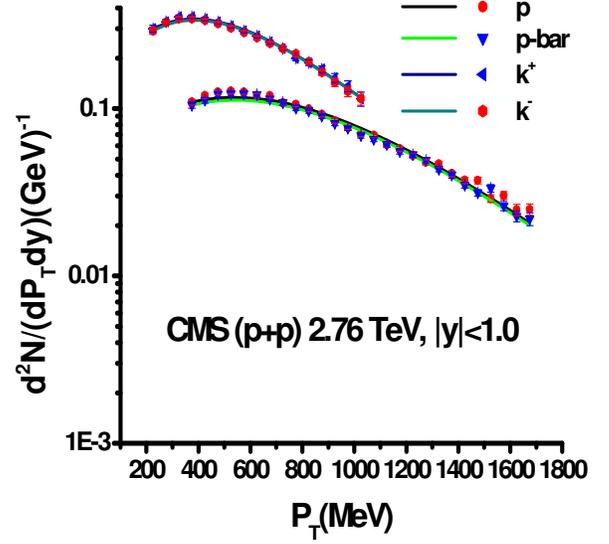

**Figure 12:** Transverse momentum spectra of p, $\bar{p}$, $K^+$ and $K^-$ at $\sqrt{s_{NN}}$ = 2.76 TeV.

From the Table 4, it is clear that the multi-strange heavier baryons freeze-out earlier than the singly strange and the other lighter non-strange particles. Also the flow parameter decreases as one goes from lighter particles to heavier multi-strange particles.

We have also successfully reproduced the transverse momentum distributions of protons and Kaons at $\sqrt{s_{NN}}$ = 2.76 TeV, as shown in figure 12. The experimental data [31] is fitted quite well with our model calculations. The various freeze-out conditions obtained from the spectra are tabulated in Table 4. The error bars here represent the sum of statistical and systematic uncertainties.

| Particle | T(MeV) | $\beta_T^0$ | n | $\chi^2$/dof |
|---|---|---|---|---|
| $p$ | 174 | 0.73 | 1.11 | 1.02 |
| $\bar{p}$ | 175 | 0.73 | 1.11 | 1.0 |
| $K^+$ | 173 | 0.70 | 1.18 | 0.41 |
| $K^-$ | 174 | 0.70 | 1.18 | 0.55 |

**Table 4.** Freeze-out conditions of protons and Kaons produced at $\sqrt{s_{NN}}$ = 2.76 TeV with corresponding $\chi^2/DoF$

### 6. Transverse momentum spectra at RHIC.

In order to compare the system properties in p-p collisions at LHC and RHIC, we have reproduced the transverse momentum distributions of p, $\bar{p}$, $K^+$ and $K^-$ produced in minimum bias p-p collisions at highest RHIC energy of $\sqrt{s_{NN}}$ = 200GeV as shown in figures 13 and 14. The experimental data has been taken from STAR experiment [32]. A good agreement is seen between our model calculations and the experimental data. The value of *n* comes out to be unity. The freeze-out conditions obtained by fitting the $p_T$ spectra of these particles are given in table 5 below.

| Particle | $T$ (MeV) | $\beta_T^0$ | $\chi^2$/dof |
|---|---|---|---|
| $p$ | 164.0 | 0.21 | 3.29 |
| $\bar{p}$ | 164.0 | 0.16 | 3.02 |
| $K^+$ | 165.0 | 0.15 | 1.89 |
| $K^-$ | 165.0 | 0.17 | 0.82 |

**Table.5 Freeze-out parameters of *p*, $\bar{p}$, $K^+$ and $K^-$ obtained from their corresponding $p_T$ spectra at $\sqrt{s_{NN}}$ = 200 GeV.**

It is obvious from Table 5 that the flow is almost insignificant in p-p collisions at RHIC. This is because of the less particle production which leads to a small system size such that the collective hydrodynamic effect at the final freeze-out is almost absent in the p-p collisions at the RHIC energies. On the other hand, the system formed in p-p collisions at LHC exhibits clear signature of a stronger hydrodynamical flow, as shown in Tables 2, 3 and 4. Thus we can conclude that there is an onset of flow in the system formed in the p-p collisions at LHC which is because of the enhanced particle production taking place at these LHC energies compared to the highest RHIC energy. Moreover, the larger values of $\chi^2$/dof obtained for protons and kaons at $\sqrt{s_{NN}}$ = 200 GeV indicates that the system is not in complete local thermal equilibrium as expected at LHC. Hence the formation of a hot and dense system attaining a reasonable degree of thermo-chemical equilibrium in a hadronic resonance gas phase before the final freeze-out takes place in p-p collisions at LHC. The system also possesses a significant collective hydrodynamic flow.

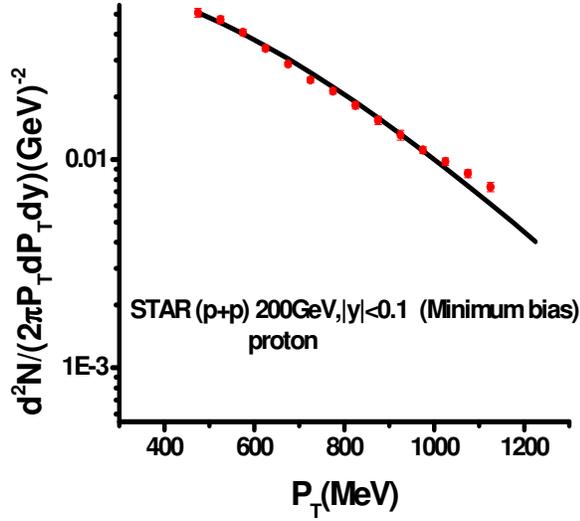
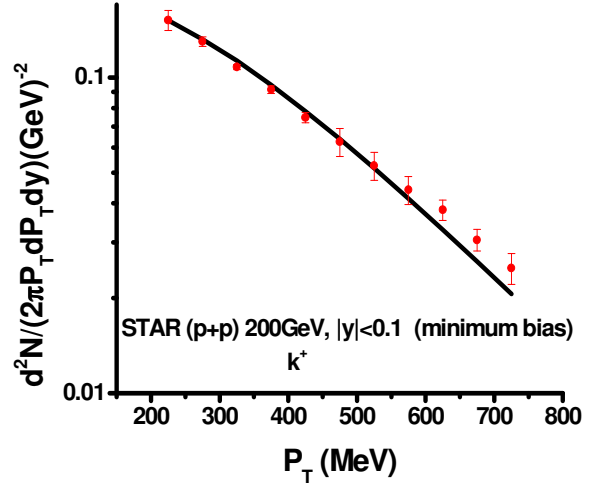
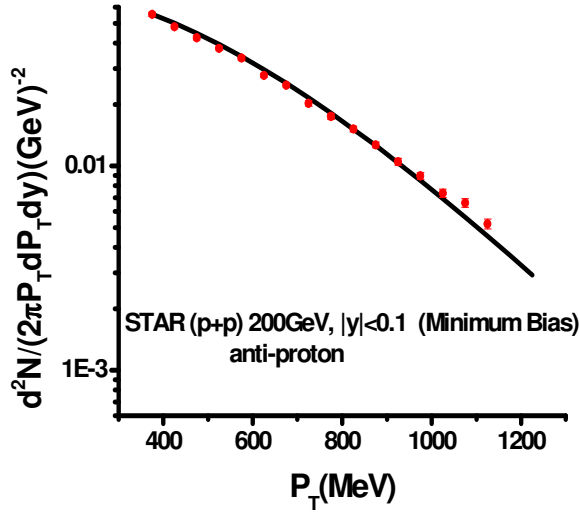
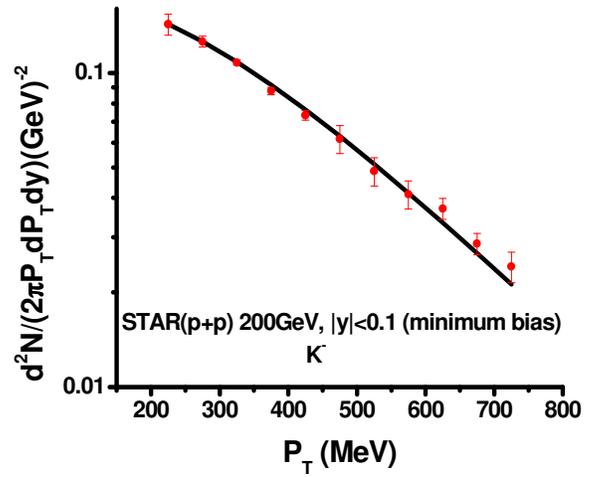

**Figure 13.** Transverse momentum spectra of protons and antiprotons in minimum bias pp collisions at 200GeV. Errors are the quadratic sum of statistical errors and point-to-point systematic errors.

**Figure.14** The $p_T$ spectra of Kaons and antiKaons in minimum bias pp collisions at 200 GeV.

In Figure 15 we present the variation of transverse flow parameter and the freeze-out temperature of protons and Kaons with the collision energy. It is seen that the flow is almost absent in p-p collisions at RHIC and increases significantly when one goes to LHC energies. At LHC, an enhancement in the transverse flow is observed on going from $\sqrt{s_{NN}}$ = 0.9 TeV to $\sqrt{s_{NN}}$ = 7.0 TeV, whereas the thermal freeze-out temperature is observed to maintain almost constant value of 173±1 MeV from $\sqrt{s_{NN}}$ = 0.9 TeV to $\sqrt{s_{NN}}$ = 7.0 TeV. Also it seems that at energies greater than 7 TeV, the freeze-out temperature may tend to decrease with a continuous increase in flow velocity, thus the system dynamics in p-p may show an even closer resemblance with the dynamics of a heavy ion collision system above these energies.

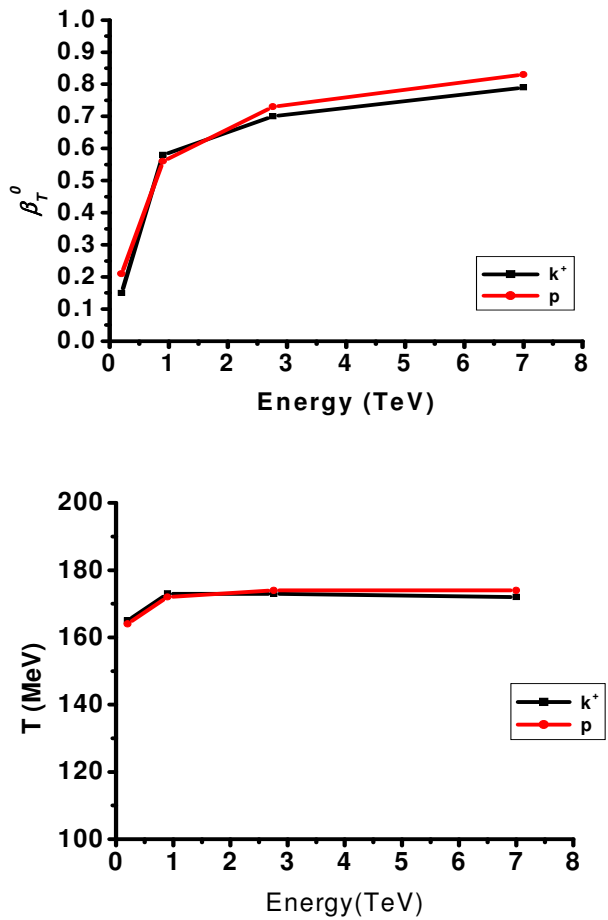

**Figure 15. Variation of flow parameter and freeze-out temperature with collision energy.**

## 7. Summary and Conclusion

The transverse momentum spectra of the hadrons (p, p̄, $K^+$, $K^-$, $K_s^0$, $\phi$, $\Lambda$, $\bar{\Lambda}$, $\Xi^-$, $\overline{\Xi^-}$, ($\Xi^-$ + $\overline{\Xi^-}$), $\Omega$, and $\bar{\Omega}$) and the rapidity distribution of the strange hadrons ($K_s^0$, ($\Lambda$ + $\bar{\Lambda}$), ($\Xi^-$ + $\overline{\Xi^+}$) at $\sqrt{s_{NN}}$ = 0.9 TeV and 7.0 TeV are fitted quite well by using our statistical thermal freeze-out model. The result extracted from the rapidity distributions of hadrons show that the chemical potential decreases with increase in the collision energy and almost vanishes at 7.0 TeV. This indicates the effects of almost complete transparency in p-p collisions at LHC. The LHC results show the existence of significant hydrodynamic flow present in the p-p system. A comparison between the transverse momentum spectra of protons and kaons at RHIC and LHC is made. For the system formed in p-p collisions at RHIC the collective behavior at final freeze-out is almost absent. We also observe an earlier freeze-out of multi-strange particles as compared to lighter mass particles. Protons and Kaons are found to freeze-out almost simultaneously at all studied energies. The earlier freeze-out of heavier hyperons is indicated by their larger thermal temperature and smaller flow parameters. The spectra are also compared with the predictions from PYTHIA and PHOJET event generators and it is found that a better fit is obtained by using our model. The variation of freeze-out conditions with energy is studied which shows an increase of both, the transverse flow and freeze-out temperature with increase in collision energy from RHIC to LHC. Also at LHC, the system is seen to attain a constant freeze-out temperature.

**Acknowledgements**

Inam-ul Bashir is thankful to the University Grants Commission (UGC) for awarding the Basic Scientific Research (BSR) Fellowship. Riyaz Ahmed Bhat is grateful to the Council of Scientific and Industrial Research (CSIR), New Delhi for awarding Senior Research Fellowship.